\documentclass{amsart}
\usepackage{graphicx}
\usepackage{epsf}

\def\noi{\noindent}
\def\be{\begin{equation}}
\def\ee{\end{equation}}
\def\dps{\displaystyle}
\def\a{\alpha}

\begin{document}

\vspace{-4truecm} {}\hfill{DSF$-$07/2006}
\vspace{1truecm}

\title[Teaching Theoretical Physics: the cases of E. Fermi
and E. Majorana]{Teaching Theoretical Physics: \\
the cases of Enrico Fermi and Ettore Majorana}%
\author{A. De Gregorio}%
\address{{\it A. De Gregorio}: Dipartimento di Fisica,
Universit\`{a} degli Studi di Roma ``La Sapienza'' \\
p.le Aldo Moro, 00185 Rome, Italy ({\rm
Alberto.DeGregorio@roma1.infn.it})}
\author{S. Esposito}%
\address{{\it S. Esposito}: Dipartimento di Scienze Fisiche,
Universit\`a di Napoli ``Federico II'' \& I.N.F.N. Sezione di
Napoli, Complesso Universitario di M. S. Angelo, Via Cinthia,
80126 Naples, Italy ({\rm Salvatore.Esposito@na.infn.it})}%


\begin{abstract}
We report on theoretical courses by Fermi and Majorana, giving
evidence of the first appearance and further development of
Quantum Mechanics teaching in Italy. On the basis of original
documents, we make a comparison between Fermi's and Majorana's
approaches. A detailed analysis is carried out of Fermi's course
on Theoretical Physics attended by Majorana in 1927-28. Three
(previously unknown) programs on advanced Physics courses
submitted by Majorana to the University of Rome between 1933 and
1936 and the course he held in Naples in 1938 complete our
analysis: Fermi's phenomenological approach resounded in Majorana,
who however combined it with a deeper theoretical approach, closer
to the modern way of presenting Quantum Mechanics.
\end{abstract}

\maketitle


\section{Introduction}

\noindent The old-dated tradition going back to Galileo Galilei
made the study of the Physical Sciences in Italy essentially meant
as {\it experimental} still in the 1920s, though the theoretical
aspect was not thoroughly excluded from the scientific debate. In
this respect, the case is emblematic of Enrico Fermi, whose thesis
for a degree in Physics at the University of Pisa, ``according to
tradition, should be in Experimental Physics;'' \cite{segrefermi}.
Then, it is not surprising that, in Italy, no stable chair of
Theoretical Physics was yet established at that time nor that the
scientific debate on the emergent novel Quantum Mechanics was
still languishing, all the more so if one compares it with the
fruitful discussions abroad.

This outline started changing with the first competition for a
stable chair of Theoretical Physics in 1926: the young Enrico
Fermi then got a position as a professor in Rome and strove to
introduce the novel concepts of Quantum Mechanics into advanced
academic courses. In the following years, a whole generation of
novel-minded physicists (both theoreticians and experimentalists)
was formed by Fermi and his coworkers. Their relevant
contributions extended to several areas of Physics, in Italy as
well as in other countries. It is then of key importance to
analyze Fermi's courses on Theoretical Physics. This analysis is
made possible by some record books reporting the content of
Fermi's lectures in Rome. Together with the textbooks on
Theoretical Physics Fermi wrote in those same years, these
documents lead to a clear and comprehensive picture of Fermi's
courses.

Most of the well versed experimentalists who worked in close
collaboration with Fermi in the 1930s had attended Fermi's
theoretical courses. On the other hand, most of the well versed
theoreticians who collaborated with Fermi's group in the Thirties,
like Gian Carlo Wick and Giovannino Gentile, had not attended
university courses in Rome. One relevant exception was Ettore
Majorana, who followed Fermi's course on Theoretical Physics
together with Segr\`e and Edoardo Amaldi in 1928.

Majorana contributed significantly to some theoretical aspects
underlying the researches performed under the supervision of Fermi
and in general by Fermi's associates. However, we will focus here
on Majorana's own progress concerning the basics (and the
applications) of Quantum Mechanics. Majorana's lectures on
Theoretical Physics very effectively give evidence of this
progress. He delivered them in 1938, when he obtained a position
as a full professor in Naples (just before he mysteriously
disappeared): that was the first chair of Theoretical Physics ever
established in Italy since the one Fermi obtained in 1927. A
recent analysis \cite{weylmajo} showed that Majorana's 1938 course
was very innovative for that time.

Before then, Majorana submitted three programs concerning advanced
Physics courses he would have held at the Rome University between
1933 and 1937. Through the recent retrieval of these programs,
whose existence was unexpected, we shall follow the evolution of
Majorana's approach to Quantum Mechanics teaching.  Further, we
shall point out how Fermi's and Majorana's courses relate to each
other, showing conceptual differences as well as impressive
similarities that denote particularly relevant teaching
strategies.


In the following section, we give an account of Fermi's lectures
on Theoretical Physics in Rome. In Sect. 3, we point out the main
stages leading Majorana to his professorship at the University of
Naples and describe the relevant features of his lectures. The
progress of Majorana's approach to Quantum Mechanics, from Fermi's
lectures to his own lectures on Theoretical Physics, is dealt with
in Sect. 4. In the subsequent section, the common traits of
Fermi's and Majorana's courses are discussed. In Sect. 6, we put
forward some speculations on the missing notes of the first four
lectures by Majorana, also discussing his teaching strategies in a
broader perspective. We give our conclusions in Sect. 7. The
programs of the three courses presented by Majorana at the
University of Rome from 1933 to 1936 are reported in the Appendix,
together with a list of the topics Fermi lectured on in his
1927-28 course in Theoretical Physics.

\section{Fermi's lectures on Theoretical Physics}

\subsection{Professor in Rome}

\noindent On December 22, 1924 the Faculty of Sciences at the
University of Rome debated about the institution of a chair of
Theoretical Physics, tenaciously promoted by Orso M. Corbino,
director of the Institute of Physics and influential politician.
The chair was formally claimed on April 27, 1925. The board of
examiners met in November 1926 and proclaimed winners Enrico
Fermi, Enrico Persico and Aldo Pontremoli, respectively. Fermi was
formally appointed as a professor of Theoretical Physics at the
Royal Institute of Physics in Rome starting from January 1, 1927.
Persico and Pontremoli were appointed at the Universities of
Florence and Milan, respectively.

With the support of Corbino, Fermi had already taught at the
University of Rome, lecturing on Mathematics to chemists and
naturalists in 1923-24; then he spent two more academic years
lecturing on Mathematical Physics in Florence. On January 20, 1927
he gave his first lecture on Theoretical Physics in
Rome\footnote{It should be noted that although Fermi's chair was
the first one ever established in Italy, his course was not the
first one ever given on Theoretical Physics. As an example, we
only mention the case of Antonio Carrelli, who regularly lectured
on Theoretical Physics in Naples since 1924.}. Noticeably, his
course was annotated by Carlo Dei and Leonardo Martinozzi and then
published \cite{fermidei}; the record book reporting the content
of Fermi's lectures is available as well, kept at the Archives of
the University of Rome ``La Sapienza.'' A strict correspondence
can be established between the topics reported in the handbook and
those that Fermi actually lectured on in 1927.

An even deeper insight into Fermi's lectures on Theoretical
Physics is provided by the book he published in the following year
\cite{fermi}. The 1927-28 teacher's record book is now available
as well, and we shall see below how strong the correspondence is
between the content of the lectures Fermi held in 1927-28 and the
book he published in 1928. For the moment, we only mention that
Ettore Majorana together with Edoardo Amaldi and Emilio Segr\`{e}
followed precisely this course. They switched from Engineering to
Physics just following the appeals by Corbino, who was striving to
build up a forefront school of Physics upon Fermi.

Amaldi moved to Physics on November 9, 1927, at least formally,
followed by Segr\`{e} on February 8, 1928 \footnote{This
information comes from the Archives of the University of Rome ``La
Sapienza", where the reports of the Board of the Faculty of
Science are kept. It is intriguing that, according to Amaldi and
Segr\`{e}, they passed from Engineering to Physics just in the
reverse order \cite{amaldi}, \cite{segre}.}; both took their exams
in Theoretical Physics in July 1928 like Majorana, though the
latter was still a student in Engineering and obtained permission
to move to Physics only on the following November 19.\footnote{We
point out that the only exam Majorana took in 1928 was that in
Theoretical Physics, according to his scholastic career report.
His mark was expressed in hundredths, as accustomed for engineers,
while Amaldi and Segr\`{e}'s in thirtieths \cite{amaldi},
\cite{segre}.} Majorana and Amaldi also enjoyed another course
held by Fermi, who also lectured on Earth Physics in 1928-29. They
took the corresponding exam on June 27, 1929, just few days before
they graduated on July 6.

\subsection{The 1927-28 course in Theoretical Physics}

\noindent An outline of the main points of the course in
Theoretical Physics held by Fermi in 1927-28 can be easily drawn
from his teacher record book. Fermi gave sixty-five lectures (the
last six ones aimed at summarizing the whole course). He started
with the Kinetic Theory of gases and the basic elements of
Statistical Mechanics: in a first set of lectures Fermi expounded
classical topics such as the dependence of pressure on the
molecular kinetic energy, mean free path, equipartition of the
energy, the phase space, the Boltzmann distribution law and the
Maxwell velocity distribution. A second set dealt with
electromagnetism: electromagnetic perturbation, the Poynting
vector, the electronic theory of dispersion, radiation theory.
Fermi then passed to introduce the foundations of the atomic
theory, lecturing on the electron, the radioactive transmutation,
Rutherford's atomic model, light quanta, the Compton effect, and a
short mention to Bohr's atom and energy levels. Therefore the
`core' of Fermi's course took shape, taking twenty lectures on
Bohr's model of the atom. Energy levels, the Rydberg constant,
stationary states, the Sommerfeld conditions, the fine structure,
the correspondence principle, selection rules, the Bohr magneton,
spatial quantization, the Zeeman and the Stark effects are some of
the topics dealt with in lectures from n. 28 to n. 47.

One striking feature emerges here: apart from a fleeting mention
of the ``quantum theory of the hydrogen atom'' in lecture n.46,
all these lectures rested on the framework of Old Quantum Theory.
The same holds for the remainder of Fermi's course, essentially
devoted to the spectra of alkali metals, alkaline earth metals and
atoms with three valence electrons.

Another trait worth remarking is that relativity was not mentioned
among the issues of Fermi's 1927-28 course. The same holds for his
1926-27 course where, although the Compton effect was explicitly
reported, the relativistic kinematics was not discussed.
Therefore, one might even wonder how Fermi could lecture on the
Compton effect at all. Remarkably, he carried out his calculations
neglecting terms of order greater than the first in the difference
$\nu - \nu^\prime$ of the light frequency and then specified:
\begin{quote}
The final formula is strictly exact. That might appear a nonsense,
since we neglected such terms as (d$\lambda)^2$. However, it
should be noted that we had already made another approximation
when we expressed the momentum of the electron as $mv$ instead of
$mv/\sqrt{1 - \beta^2}$ therefore neglecting relativity. Now, the
curious affair is that an opposite error creeps in this way, so
that the formula above turns out to be strictly exact instead of
being an approximation (see Ref. \cite{fermidei}, pp. 81-82).
\end{quote}

The strict correspondence between the topics reported in the
1927-28 record book and those explained in the 1928 book in Ref.
\cite{fermi} can be readily realized by merely comparing the
content of the first nine lectures ranging from November 15 to
December 3, 1927 with the titles of the sections in the first
chapter of Fermi's book. We give both in Table 1. The whole of the
contents of Fermi's second course on Theoretical Physics is
instead reported in the Appendix. Given such a correspondence
between the lectures and the book in Ref. \cite{fermi}, we can
group the lectures into six sets: each relates to one of the first
six chapters of Ref. \cite{fermi} as illustrated in Table 2. The
correspondence is very strict in the case of the first two
chapters, while slight differences come out starting from the
third chapter: sometimes the order of the topics of the course is
altered in comparison with the book, and in the latter several
arguments might receive a somewhat more detailed treatment.

\begin{table}
\begin{center} \small
\begin{tabular}{|ll|ll|}
\hline \hline & Fermi course (1927-28) & & Fermi book (1928) \\
\hline \hline
I)    & Interactions between molecules. & 1. & Molecular
interactions. \\
      & Solids and fluids in the framework & & \\
      & of the kinetic theory.  & & \\
II)   & Dependence of pressure upon & 2. & Molecular mobility in solids \\
      & molecular kinetic energy. & & and fluids. \\
III)  & Mean free path. Vapor rays. & & \\
IV)   & Kinetic energy in general coordinates. & 3. & Kinetic theory of ideal gases. \\
      & The theorem of equipartition & & \\
      & of energy (without proof). & 4. & Mean free path. \\
V)    & Phase space. & & \\
VI)   & Evaluation of the probability of & 5. & Equipartition of energy. \\
      & distribution. & & \\
VII)  & The Boltzmann distribution law. & 6. & Boltzmann statistical \\
VIII) & Proof of the principle of & & distribution. \\
      & equipartition of the energy. & & \\
IX)   & The Maxwell law. & 7. & The Maxwell law. \\ \hline \hline
\end{tabular}
\end{center}
\caption{A comparison between the topics covered in the first nine
lectures of Fermi's 1927-28 course, and the sections in the first
chapter of his book (Ref. \cite{fermi}).}
\end{table}

\begin{table}
\begin{center} \small
\begin{tabular}{|ll|ll|}
\hline \hline & $\hspace{-0.7truecm}$Sets of lectures & &
$\hspace{1.2truecm}$ Chapters \\ \hline \hline
n. \hphantom{0}1 - n. \hphantom{0}9:   & Nov. 15 - Dec. 3 & I &
Kinetic theory of gases \\ \hline %
n. 10 - n. 19: & Dec. 6 - Jan. 14 & II  &
Electromagnetic theory of light \\ \hline %
n. 20 - n. 23: & Jan. 17 - Jan. 24            & III & Electric
particles \\ \hline %
n. 24 - n. 27: & Jan. 26 - Feb. 2             & IV  & Energy
exchange between \\
& & & light and matter \\ \hline %
n. 28 - n. 47: & Feb. 4 - Mar. 31             & V   & Bohr's atom \\ \hline %
n. 48 - n. 59:  & Apr. 17 - May 15            & VI  &
Spectral multiplicity \\ \hline %
n. 60 - n. 65: & May 19 - Jun. 2 & & \\
            & (summary of the course) & &  \\ \hline \hline
\end{tabular} \end{center}
\caption{Correspondence between the lectures of the course on
Theoretical Physics Fermi held in 1927-28 and the first six
chapters of his book \cite{fermi}.}
\end{table}

\section{Ettore Majorana: from Fermi's courses to professorship}

\noindent Ettore Majorana began to frequent the Institute of
Physics at the University of Rome at the end of 1927 or in the
first months of the following year \cite{amaldi}, \cite{segre},
\cite{meeting}, \cite{dig}. It is very likely that the first
course he attended at the Institute of Physics was precisely the
one in Theoretical Physics given by Fermi and discussed in the
previous section. Some indirect records of this can be traced in
Majorana's ``{\it Volumetti}'' \cite{volumetti} (``Booklets''),
which also contain several issues probably discussed by Fermi in
his course on Earth Physics. Since no lecture notes of Fermi's
courses followed by Majorana are known, we can only compare what
is reported in the {\it Volumetti} with Fermi's book \cite{fermi}.
In this respect, it is noteworthy that the approach to Quantum
Mechanics contained in the {\it Volumetti} substantially differed
from Fermi's one. That denotes an autonomous work by the student
Majorana, who was particularly impressed by the book on Group
Theory and Quantum Mechanics \cite{Weyl} Hermann Weyl published at
the end of 1928 (see the analysis in Ref. \cite{weylmajo}).
Nevertheless, the more phenomenological approach of Fermi was
always latent and never abandoned by Majorana, as we shall see
later. In those years, Majorana aimed at a full understanding of
the novel Mechanics, based on solid and general mathematical
grounds but without disregarding evidence coming from atomic and
nuclear experiments.

\subsection{Majorana's involvement with Physics teaching}

\noindent Besides Majorana's {\it Volumetti}, his ``{\it
Quaderni}'' (``Notebooks''), whose analysis has just began, record
research studies up to approximately 1932 (see \cite{volumetti}).
In those years, Majorana was used to frequent the Fermi group in
Rome, although without even seeking for a stable position, which
he would only obtain in Naples at the end of 1937. However,
Majorana succeeded in obtaining the professorship degree of {\it
libero docente} (an academic title analogous to the German {\it
Privatdozent}) in January 1933 and, due to this position, he
proposed some academic courses. The programs concerning three
courses he would have given between 1933 and 1937 have only
recently been discovered in the Archives of the University of
Rome. This retrieval is very important and surprising, since these
original documents cover a period that has been referred to as
Majorana's gloomy years by the testimonies of that epoch (see, for
example, \cite{amaldi}, \cite{segre}, \cite{recami}), during which
he secluded himself from the life of the Institute.

The three programs are here translated in Appendixes B, C, D. Even
at a first look, it is evident that Majorana was very careful in
choosing the topics he would have treated in his proposed courses,
and this reveals his genuine interest in advanced physics
teaching.

In 1933, Majorana benefited from a C.N.R. visiting fellowship at
the Institute of Theoretical Physics in Leipzig directed by
Heisenberg. He went back to Italy for the Easter holidays between
the end of April and the beginning of May, and his first program
was submitted precisely at this time: the novel and stimulating
atmosphere found in Leipzig probably had spurred him. He
reiterated his request to give advanced Physics courses also in
subsequent years, but he never delivered any course in Rome. This
is implicitly testified by his friends and colleagues and
explicitly confirmed by Majorana himself, who in the three
programs declared he had not given any course before: a
circumstance likely due to the lack of students.

The first two programs correspond to the courses in Mathematical
Methods of Quantum Mechanics and Atomic Physics, respectively, and
partially overlap. In presenting Quantum Mechanics, Majorana aimed
to use group-theoretic methods: that was an unusual approach for
the times, as already discussed \cite{weylmajo}, and will again be
adopted in Majorana's lectures at the University of Naples in
1938. The approach will become the standard one in the teaching of
Quantum Mechanics only many years after the Second World War.

The second program contains references to questions more
phenomenological in nature and the third one deals with Quantum
Electrodynamics: that was again an unusual topic for the Italian
academic courses of the period, but had been fascinating Majorana
for a long time, as clearly emerges from the unpublished research
notes in his {\it Quaderni}.

\subsection{Lecturing at the University of Naples}

\noindent In 1937, the national competition for a full
professorship in Theoretical Physics at the University of Palermo
provided a good chance for entering advanced Physics teaching, for
Majorana as well as for other skilful young physicists such as
Gentile, Wick, Leo Pincherle, Giulio Racah, and Gleb Wataghin.
Majorana was urged to take part in the competition by his
colleagues and friends \cite{amaldi}, \cite{recami}. In
particular, Fermi persuaded him to finally publish an earlier work
of his, on the symmetrical theory of electrons and positrons
\cite{symmetrical}, which will become famous for the appearance of
the so-called `Majorana neutrino' theory. Majorana had already
faced (and solved) this problem long before, as clearly appears in
his {\it Quaderni}. An explicit mention of ``the positive electron
and the symmetry of charges'' was also recorded in the third
program, which Majorana had submitted on April 28, 1936.

Without going through the details of the 1937 competition (already
discussed in the literature \cite{amaldi}, \cite{segre},
\cite{recami}), we just recall that Majorana was eventually
appointed as a full professor of Theoretical Physics at the
University of Naples ``for high and well-deserved repute,
independently of the competition rules.'' He gave only twenty-one
lectures at the Institute od Physics in Naples, from January 13 to
March 24, before his mysterious disappearance on March 26, 1938
\cite{bibliopolis}, \cite{moreno}.

Fortunately, some manuscripts of the lecture notes written by
Majorana himself for the students of his course on Theoretical
Physics are available. He entrusted them to one of his students,
Ms. Gilda Senatore, before he disappeared. They are deposited at
the Domus Galil{\ae}ana in Pisa and were reproduced in a book some
years ago \cite{bibliopolis}. However, a transcription comprising
six lectures not included in the Pisa collection has recently been
discovered \cite{moreno}. Therefore, in order to go through
Majorana's teaching strategies, the analysis of his course should
be updated (see \cite{espobiblio}) and embrace some previously
unknown topics, mainly Relativity (an account of this is in
\cite{weylmajo}).

Even if a comprehensive record of the course Majorana actually
held in Naples in 1938 is not available, some peculiar -
historical and scientific - features can be drawn from his very
lecture notes.

In the prolusion, delivered on January 13, Majorana gave a broad
outline of his course, aimed at the study of Quantum Mechanics and
its applications to Atomic Physics (see Refs. \cite{bibliopolis},
\cite{weylmajo} and \cite{espobiblio}). He also stated his
teaching method, which would in fact consist of a combination of
two: he mentioned the ``mathematical method'', through which ``the
quantum formalism is presented in its most general and therefore
its clearest propositions from the very beginning and only
afterwards are the criteria for applying it explained;'' and he
mentioned what he called the ``historical method,'' which
``explains how the first idea of the formalism appeared.''
Majorana regarded them as two opposite methods, but he aimed at a
fruitful synergy between them.

It is really intriguing to note how closely Majorana's ideas
resemble the preface to Paul A.M. Dirac's book on ``The Principles
of Quantum Mechanics.'' As for the mathematical presentation of
quantum theory, Dirac wrote in 1930: ''The symbolic method [...]
deals directly in an abstract way with the quantities of
fundamental importance,'' allowing the physical laws to be
expressed ``in a neat and concise way;'' as for the applications,
they ``follow strictly from the general assumptions.'' However,
all this implied ``a complete break from the historical line of
development'' \cite{dirac}.

We do not have any written notes of Majorana's lectures from the
second to the fifth ones and we will speculate on their possible
content in Section 5.

The remainder of the course is approximately divided into three
parts:
\begin{itemize}
\item[-] Old Quantum Mechanics (3+1/2 lectures); %
\item[-] Electromagnetism, Relativity and their application to
microscopic phenomena (6+1/2 lectures); %
\item[-] Mathematical foundations of Quantum Mechanics (6
lectures).
\end{itemize}
In the first of these parts Majorana is probably expanding on
topics introduced in the first four lectures. He deals with the
relevant phenomenology of Atomic Physics and its interpretation in
the framework of the Old Quantum Theory of Bohr and Sommerfeld. In
particular, starting from the available spectroscopic data, he
introduces the spin and the related Pauli exclusion principle,
accounting for the periodic table of the elements and the features
of the spectra of one- and two-electron atoms in the Sommerfeld
theory.

The second part starts with the classical radiation theory,
reporting explicit solutions of the Maxwell equations for a given
system; some applications (e.g. scattering of light on the
atmosphere) are discussed as well. Then Majorana deals with the
Theory of Relativity, starting from simple phenomenology and later
introducing the appropriate mathematical formalism. Lorentz
transformations along with their immediate consequences are
introduced in a simple and original way, and applications to the
electromagnetic field are extensively deduced. Particular emphasis
is given to the relativistic dynamics of electrons, which Majorana
obtains from a variational principle.

The effort gone into Electromagnetism and Relativity ends up with
the discussion of processes like the photoelectric effect, Thomson
scattering and the Compton effect. The Franck {\&} Hertz
experiment is also treated here, exhibiting the existence of
energy levels. All these phenomena are considered as the
phenomenological bases for Quantum Mechanics, and this appears
quite `in contrast' with the approach of the first part of the
course, based on spectroscopic phenomenology.

The third set of lecture notes is more mathematical in nature. The
(non-relativistic) wave equation along with its statistical
interpretation based on the Heisenberg uncertainty principle are
presented only after a detailed discussion of some relevant topics
on matrix and operator theory and Fourier transforms.

The original manuscripts collecting Majorana's lecture notes
include a last discorsive part (see \cite{bibliopolis},
\cite{path}, \cite{espobiblio}). It has been primarily interpreted
as the notes of a lecture that Majorana would have held at the
University of Naples if he did not mysteriously disappear.
However, as already noted in \cite{path}, an accurate analysis of
the content of this manuscript seems not to confirm a similar
conclusion. In fact, the topics are rather advanced and do not
relate, to any extent, to the last lecture (the 21st one) on the
uncertainty principle. Moreover, the style of presentation of the
topics is completely different from that used in the main body of
the Naples lectures. Rather, the expository form resembles the
style he had used in his prolusion.

\section{The progress of Majorana's approach to Theoretical Physics}

\noindent Majorana's approach to Theoretical Physics is well
traced in his {\it ``Volumetti''} \cite{volumetti}. Standard
mathematical tools (differential equations, Fourier integrals and
so on) dominate approximately the first half of these documents.
They are used in order to solve well-defined physical problems,
such as, for example, the heat propagation in an isotropic and
homogeneous medium or the equilibrium of a rotating fluid
(Clairaut's problem), or even to face contemporary problems such
the scattering of an $\alpha$ particle by a radioactive nucleus or
the solution of the Thomas-Fermi equation with the appropriate
boundary conditions for atoms \cite{fermithomas}, \cite{dig}.

Majorana's approach, however, sensibly changes in the remainder of
his {\it ``Volumetti''}, dating to the end of 1928 and largely
devoted to Group Theory and symmetries, which topics underlie
almost all his articles. In this respect, an illuminating example
is that of the infinite-dimensional unitary representations of the
Lorentz group \cite{weylmajo}, whose study, reported in detail in
the {\it Volumetto V}, forms the basis for his important article
of 1932 \cite{infinite} on a relativistic theory of particles with
arbitrary spin.

An approach similar to the latter is also present in the three
proposed programs, but some comments are needed here.

From what reported in Appendix B, it is evident that the program
on Mathematical Methods of Quantum Mechanics submitted in 1933 is
{\it entirely} based on group-theoretic methods. Majorana mentions
the elements of unitary geometry and transformations, pointing at
the modifications to classical kinematics and at the invariance
under transformation group. Then he goes through the rotation and
permutation groups. The `standard' presentation of Quantum
Mechanics (both in the Schr\"odinger version and in the Heisenberg
one) is completely removed from this program. Even the
applications (e.g. angular momenta, systems with identical
particles) are strictly related to the formalism of Group Theory.
The same applies to the Theory of Relativity, mentioned in the
framework of the Lorentz group and of spinor calculus.

Majorana's attitude towards transformation theory resembles what
Dirac wrote in his preface about the formulation of the
fundamental laws of Nature, which ``requires the use of the
mathematics of transformations. The important things in the world
appear as the invariants [...] of these transformations.''
Further, almost all the topics mentioned in Majorana's 1933-34
program were explicitly treated by Weyl in his book: the item 1 in
the program appears in the chapter 1 of Ref. \cite{Weyl}, entitled
precisely ``Unitary Geometry''; the items from 3 (partially) to 7
of Majorana, mentioning representations and applications of Group
Theory to momenta, identical particles, and relativistic
particles, are also dealt with in chapters 3, 4 and 5 of the Weyl
book. An exception seems to be the discussion of the phase space
and the introduction of the quantum of action in item 2, which
does not appear to correspond to any topic in Weyl's book.

The proposed impressive use of Group Theory is certainly justified
by Majorana's own attention to conciseness and generality, which
is constantly present in his studies. We saw that Majorana himself
explicitly stated this attention in the prolusion to his course,
where he also followed Dirac. The latter gave up the usual
``method of co-ordinates or representations'' through his own
preference for the ``symbolic method.'' However, also the line of
thinking followed by the Leipzig school and experienced by
Majorana in 1933 \cite{amaldi}, \cite{recami}, just when he
officially submitted his first program, may have played some role
on Majorana's attitude to Group Theory.

Things are somewhat different in the 1935-36 program on
Mathematical Methods of Atomic Physics (we recall no program is
available relative to 1934-35). A more phenomenological approach
emerges, compared with that proposed in 1933-34. The
group-theoretic methods appear only marginal and subordinated to
practical applications, like complex atomic spectra and hyperfine
structures. It would seem quite evident that the choice of the
items for this course was influenced by the line of thinking
adopted by the Rome group headed by Fermi, who had been engaged in
spectroscopic researches before switching to nuclear physics
\cite{radio}. Further, in the program submitted by Majorana, the
mention of the elements of Nuclear Physics precisely follows that
of hyperfine structures. This sequence is in agreement with the
course of events recalled by Segr\`e \cite{segre} in giving
account of the switching from atomic to nuclear researches in
Rome.\footnote{Note also that a contribution of Majorana is
acknowledged in the paper by Fermi and Segr\`e on hyperfine
structures \cite{iperfine}.}

On the contrary, the 1936-37 program of Quantum Electrodynamics
appears free from external influences, probably due to the absence
of Majorana from the Institute of Physics in those years
\cite{amaldi}. The program is largely based on personal studies
which are easily recognizable in his handwritten notes (the {\it
``Quaderni''} and partially the {\it ``Volumetti''}).

A further elaboration by Majorana of his own teaching approach
comes out when we compare the known lecture notes of the course he
held in Naples with the three programs submitted in Rome. The
phenomenological part of Majorana's 1938 course strictly resembles
Fermi's 1927-28 course. The mathematical part, instead, has
contact points with some items Majorana mentioned in his three
programs. In particular, the unitary geometry (only loosely
borrowed from the Weyl book) mentioned in the first program is
then discussed in lecture n.17 of the Naples course; matrix
calculus (second program) is then introduced in lecture n.16, and
the relativistic theory of the electron (second and third
programs) is expounded in lectures n.13 and n.14 \cite{moreno}.
This mathematical part appears as peculiar of Majorana's own
teaching approach, and inspired by Weyl and the group-theoretic
methods as well as by Dirac. It is quite distant from other
contemporary approaches as, for example, those followed by E.
Persico \cite{persico}, whose book was all the same recommended to
Majorana's students \cite{moreno}.

\section{Common traits in Fermi's and Majorana's courses}

\noindent As is well-known, Old Quantum Theory essentially
superimposes few simple rules on the conceptual scheme of
Classical Mechanics. Its framework survived as a relevant tool in
Theoretical Physics courses for many years, an intuitive
representation of physical phenomena being preserved by it. It was
only about the 1950s that Quantum Mechanics could thoroughly get
in at the university courses, mainly with its Wave-Mechanical
version.

Fermi dealt with {\it The New Quantum Mechanics} in the final
chapter of his 1928 book. However, he did not lecture on Wave
Mechanics at all before his 1928-29 course in Theoretical Physics.
In the following academic year, 1929-30, Wave Mechanics took a
total of twelve lectures out of fifty-eight, comprising a brief
mention of Perturbation Theory (two lectures)\footnote{Fermi's
record book of his 1928-29 course in Theoretical Physics is
missing. However, we can infer that he devoted some lectures to
Wave Mechanics since he asked about the Schr\"odinger equation of
a candidate during an exam in October 1929 (Archives of the
University of Rome ''La Sapienza''). As for Fermi's record book of
his 1929-30 course, it should be noted that it is kept in Pisa
rather than in the Rome Archives \cite{leone}.}

The Bohr and Sommerfeld quantization rules underpin also the first
group of Majorana's lectures, up to February 3, 1938.
Furthermore, Majorana went into Matrix Mechanics in Naples since
March 8, 1938 (and even before it in the three programs he
submitted in Rome); he also lectured on Wave Mechanics starting
from March 17 until he disappeared.

We are now going through Majorana's way of treating atomic physics
in the framework of Old Quantum Theory. The legacy of Fermi's
course will be readily disclosed in Majorana's lectures by our
analysis.

\subsection{Phenomenological approach and theoretical analysis}

\noi At a first sight, Fermi and Majorana's approaches appear not
easily related to each other. In fact, a detailed relativistic
treatment of the fine structure in the hydrogen spectrum is
reported in Majorana's lectures, whereas Fermi only gave the final
formula along with a few comments.

As for Fermi's way to the fine structure, he firstly wrote down
the Newton equations for one electron in the Coulomb field of the
nucleus and then put forward the conditions to fulfill in order to
fit the spectroscopic terms. In other words, he attained the
Bohr-Sommerfeld conditions as a generalization of empirical rules.
He also dealt with the motion of the nucleus, recalling the slight
difference between the even lines of the Pickering series of
$He^+$ and the Balmer series of $H$. Fermi then remarked that a
relativistic correction to the Bohr formula should be taken into
account. Indeed, Old Quantum Theory predicted $n$ different values
ranging from $1$ to $n$ for the azimuth $k$ of an electron in the
$n$ orbit. The value $k = 1$ corresponded to the highest
eccentricity, while $k = n$ to circular orbits. Now, higher
eccentricity meant a closer approach to the nucleus and higher
velocities at the perihelion. Therefore, the lower $k$ the larger
the correction according to Old Quantum Theory. The final formula
was: %
\be %
W_{n,k} = - \frac{2 \pi^2 m Z^2 e^4}{h^2 n^2} - \frac{\pi^4 m Z^4
e^8}{c^2 h^4 n^3} \left( \frac{8}{\dps k + \frac{1}{2}} -
\frac{6}{n} \right)
\ee %
Fermi stressed that, in the framework of Quantum Mechanics (where
the azimuthal quantum number $k$ ranged from $0$ to $n - 1$), the
relativistic correction was different from the previous formula,
but the additional spin correction made the old result still hold
(we find here another case where two opposite corrections
cancelled out each other, besides the case already mentioned of
the Compton effect). Fermi expounded these topics in some
paragraphs of the fifth chapter of his {\it Introduzione}. Their
titles are as follows:
\begin{itemize}
\item[-] Motion of the electron in the hydrogen atom; %
\item[-] Hydrogen spectrum; %
\item[-] Hydrogen energy levels; %
\item[-] Spectrum of the ionized helium; %
\item[-] The Sommerfeld condition; %
\item[-] Hydrogen elliptic orbits.
\end{itemize}

They correspond to the lectures from the 28th (February 4) to the
34th (February 28), and to the 37th (March 8, 1928).

Summing up, we can say that Fermi's approach to atomic levels was
`from below', relying on mere phenomenological (Balmer series) and
intuitive (velocity of the perihelion) arguments. In this respect,
we may note that Fermi's intuitive and phenomenological approach
appears as opposed to the more abstract approach theorized for
example by Dirac, who wrote: ''A book on the new physics, if not
purely descriptive of experimental work, must be essentially
mathematical.''

Differently from Fermi, a very general method was used by Majorana
in his known lectures to calculate the correction leading to the
fine structure formula. Majorana wrote straight down the
relativistic Hamiltonian and the Hamilton equations of motion for
the electron in the hydrogen atom. Then he introduced the
Sommerfeld conditions, leading in the end to the same formula
written down by Fermi (just in a slightly different form):
\be \label{6.5} %
\nu_{n,\ell} = - \frac{W_{n,\ell}}{h} = Z^2 \frac{R}{n^2} + Z^4
\a^2 \frac{R}{n^3} \left( \frac{1}{\ell} - \frac{3}{4n} \right) .
\ee %
There is no echo of Fermi's intuitive and phenomenological
approach. However, the crucial point is that Majorana's treatment
of the relativistic corrections was {\it not} the way Majorana
introduced the quantized atom. In his notes there are several
references calling up various points that Majorana tackled in the
four missing lectures, held before January 25, 1938. For example:
``As we remember,'' the non relativistic expression for the radial
momentum is analogous to the relativistic one; furthermore, ``we
already know'' the value of the integral and the total quantum $n
= n_r$ is introduced ``here as well.'' Majorana also mentions the
$He^+$ spectrum, in such a way that we may infer he had already
dealt with it in the very first part of his course. In other
words, it is very likely that Majorana introduced such topics as
the Balmer series, the Sommerfeld conditions, and the elliptic
orbits in his first four lectures, whose notes are not available.
As a consequence, there is no evidence that his approach to the
atom was substantially different from Fermi's one. All we can say
is that Majorana dealt with the relativistic corrections in more
details than Fermi, so confirming his attitude towards
generalization. Although he roughly condensed the topics of the
whole of Fermi's course in about ten lectures, this circumstance
did not hinder him from providing a deep insight into Old Quantum
Theory.

Once the relativistic correction was given, Majorana drew his
attention to the motion of the nucleus (and differently from Fermi
he also provided the formula for the reduced mass). He pointed out
that the Rydberg constant could now agree with the value ``already
given.'' However, the empirical selection rules lacked of an
underlying theoretical account, which Majorana characterized as
one of the ``difficulties'' of the Sommerfeld theory of the atom;
note that the selection principle and the mean life of the quantum
states are further topics which follow the fine structure one in
Fermi's book (Fermi also dealt with them in his 42th lecture on
March 20, 1928).

\subsection{A clear legacy}

\noi The strong analogy with Fermi's approach to the atom becomes
more evident when Majorana deals with spatial quantization, alkali
spectra, and spin. He shows that the degeneracy of the orbits can
be resolved by applying a magnetic field to the atomic system. He
still moves within the framework of what he himself called the
``old theory,'' and he again uses the ``elementary theory'' in
dealing with the Bohr magneton. The final formula for the splitted
lines is explicitly reported:
\be %
\nu = \frac{W(n,\ell,m) - W(n',\ell',m')}{h} = \frac{W(n,\ell) -
W(n',\ell')}{h} + (m - m') \, \frac{\mu_0 H}{h}.
\ee %
It is almost identical with the formula of the Zeeman effect
reported by Fermi (Eq. (57) on page 179 of Ref. \cite{fermi}). In
fact, in the fifth chapter of his book Fermi also went into
spatial quantization, Bohr magneton, the spinning electron and, in
the end, the Zeeman effect and the Stark effect - corresponding to
lectures from the 43th (March 22) to the 47th (March 31). Going
back to Fermi's attitude for intuitive models, we also note that
Fermi introduces the Lorentz theory of the Zeeman effect before
the quantum treatment of the latter: although the early theory
``has been replaced by the quantum theory [...] it accounts for
many phenomena in a very simple way.'' Majorana instead does not
mention the Lorentz theory nor the Stark effect, which one might
ascribe perhaps to his already recalled need for brevity.

Fermi's preference for intuition and simplicity is clearly stated
at the head of the sixth chapter, devoted to spectral
multiplicity: ``We are going to exploit the Sommerfeld method to
attain a more simple and intuitive picture of the atom.'' Fermi
puts forward the distance-dependent shielding effect due to the
inner electrons, and provides an explicit expression for the
effective charge acting on the optical electron closely resembling
the one later reported by Majorana. Here again, the smaller the
azimuth, the closer the approach of the semiclassical orbit to the
nucleus at the perihelion, the stronger the attractive force on
the optical electron. Fermi provides some further details about
the terms in a non Newtonian central field, illustrating the
energy levels of sodium with a scheme. Now, the same scheme with
the $Na$ spectrum and the same phenomenological arguments
accounting for the splitting of the lines are reported also by
Majorana in his course.

Majorana's lecture of January 29, 1938 deals with Pauli's
exclusion principle, giving an account of the periodic table, and
the alkali metal spectra. The same topics were (more extensively)
reported in the sixth chapter of Fermi's book, in the following
paragraphs:
\begin{itemize}
\item[-] Central motion in a non Newtonian central field; %
\item[-] The series of the spectral terms; %
\item[-] The spectra of the alkali metals; %
\item[-] The inner quantum; %
\item[-] The counting of the quantum states; %
\item[-] The exclusion principle and the electronic rings; %
\item[-] Chemical and spectroscopic applications.
\end{itemize}
These paragraphs roughly correspond to Fermi's lectures from the
48th (April 17) to the 54th (May 3).

Majorana's notes resemble the part of Fermi's course just
mentioned not only for its general formulation, but also for many
formal details. For example, they both report many common
formulae. Fermi puts forward the deep similarity between the
electronic configurations of alkali metals, and also of some
analogous metals. As for the latter, Fermi explicitly mentions
only the case of $Cu$, $Ag$, and $Au$, as also Majorana does.
Furthermore, in dealing with the Stoner table, Fermi uses the
symbol $Cp$ for Cassiopeium, which at the time was the German
version of the French-inspired name Lutetium for the element 71;
Majorana, with the same notation, reports the symbol $Cp$ in his
Stoner table.

The most humble but, at the same time, the most eloquent example
of similarity occurs with the exclusion principle: Fermi writes
that ``at most only one electron'' exists with the given four
quantum numbers. In an early version Majorana wrote that there is
``only one electron'' in one orbit, given the four quantum
numbers. However, he then corrects his sentence specifying that
there is ``at most only one electron,'' exactly the same
expression Fermi used. One may regard this as an emblem when
describing the legacy of Fermi's course in Majorana's lectures.
However, it should be kept in mind that this legacy does not
concern only formal details but appears to involve also meaningful
aspects such as the attention paid to phenomenological
descriptions.

Once such strong correspondence is established, it is
straightforward to find some other examples of it in the first
part of Majorana's lecture devoted to {\it The spectra of the
atoms with two valence electrons} (which concludes the first set
of lectures we mentioned in the paragraph 3.2). It corresponds to
Fermi's paragraph which follows the applications of Stoner's
table, and to Fermi's lectures NN.55-56 (May 5-8, 1928). One more
topic common to both Fermi and Majorana's courses is the
scattering of light from the atmosphere.

On Thursday February 17, 1938 Majorana dealt with further issues
in the framework of Old Quantum Theory, lecturing on the Compton
effect and the Franck \& Hertz experiment. We will report on them
soon.

\subsection{Majorana's first four lectures}

\noi Neither Majorana's personal notes nor their transcription
cover the first lectures up to January 22, 1938. All the same, the
very analysis of those notes brings out what the four lectures
from January 15 to 22 should be concerned with. Majorana himself
explicitly recalled some of the topics he had already dealt with
in his first lectures: we have already mentioned the Balmer
series, the Sommerfeld conditions, the elliptic orbits. The case
can be readily added of the nomenclature of the energy levels: the
meaning of the $s$, $p$, $d$, $f$, ... levels was taken for given
when Majorana dealt with the alkali metals and their spectra.

But what about the points which Majorana might deal with before
January 25 without explicitly mentioning them in his subsequent
notes? To this purpose, we can exploit the analogy with Fermi's
course. In other words, if we take for granted that Majorana's
course followed the main lines of Fermi's lectures for what
concerned Old Quantum Theory, we can speculate about the points
dealt with by Majorana from January 15 to January 22, 1938 by
merely looking at the topics of the first chapters of Fermi's {\it
Introduzione}.

Thus, some of the first topics tackled by Majorana might well be
the kinetic theory of gases and statistical mechanics: they
introduced Fermi's course into atomic models. Fermi also showed
how the quantized levels of the harmonic oscillator could lead to
the Planck formula. Here we have two more topics Majorana should
deal with: a) As for the Planck theory, Majorana wrote down the
Planck constant in his formulae but never made a word of its
meaning nor of its value in his notes following January 22. b) As
for the harmonic oscillator, it is worth noting that the zero
point energy had been repeatedly touched by Fermi in his papers
preluding to the quantum statistics \cite{cordella}. In the first
chapter of his {\it Introduzione}, Fermi, in spite of being in the
framework of Old Quantum Theory, remarked that the zero point
energy provided a better agreement with the experiments [p. 129].
Now, it could be likely that Fermi actually mentioned the zero
point energy in his course as well, and therefore that Majorana
too touched that point, even if in dealing with Old Quantum
Theory.

The Bohr-Sommerfeld theory, the Zeeman and the Stark effects, the
selection rules and the correspondence principle are some more
topics which Majorana likely borrowed from Fermi's course in his
first four lectures.

\section{A step ahead}

\subsection{About the missing notes}

\noi The question now arises: which were the fates of the notes
concerning the four lectures from January 15 to January 22? It
appears hard to answer, but a closer look reveals something more:
maybe it is not a well posed question. The point is intriguing.

We know that on January 22, 1938 Majorana wrote from Naples to his
mother in Rome: ``I believe I will be there within a few days, but
only for few hours, since I have to take a book from Treves [it
was the name of a publishing house] and some more from home'' (the
text of the letter is reported in Ref. \cite{recami}). We recall
that the notes we have begin with Majorana's lecture of Tuesday,
January 25, 1938, the first day available after Saturday 22. Which
were the books Majorana urgently needed to bring to Naples in
those same days when he was giving his course? Is it by chance
that his notes start only after the journey he announced to his
mother? We suggest a possible outline: the course of 1938 was the
first one Majorana had ever held. After a few lectures introducing
Old Quantum Theory, he probably realized that some notes would be
useful for his student. Some textbooks on Theoretical Physics
would help him to write the notes down. Fermi's book would fit
extremely well for what concerned Old Quantum Theory, since
Majorana was very familiar with Fermi's 1927-28 course. Thus, our
tentative answer is that the notes on the first four lectures,
from January 15 to 22, had no fate at all, simply because they had
never been written down by Majorana.

An indirect check of the outline we have just suggested is as
follows. As already mentioned, Majorana came back to the motion of
the nucleus ``in the framework of non relativistic mechanics.''
That sounds like a step backwards with respect to the treatment of
relativistic corrections he had just concluded. However, his
choice seems to fit very well with our idea that Majorana had just
taken Fermi's book from Rome: maybe he had just realized that an
interesting topic was left out in his previous lectures.
\footnote{The latter is not the only episode. Take the Franck \&
Hertz experiment, which follows the Compton effect and precedes
Matrix and Wave Mechanics in Majorana's notes: though the Compton
effect appears as a direct application of the electromagnetism and
the quantum theory of light, the experiment revealing the energy
levels of mercury vapors appears rather out of context. Before the
new transcription of the notes was discovered, the Franck \& Hertz
experiment was considered part of ``a series of relevant physical
phenomena'' which Majorana tackled ``in order to put forward
masses of physical facts and results in the simplest way [...]
before he started the hard journey towards the new mechanics''
(see N. Cabibbo, in Ref. \cite{bibliopolis} pp. 118-119). Now we
know six more lectures and we have to account for the fact that
the Franck \& Hertz experiment immediately follows a whole set of
lectures on electromagnetism and relativity.}

\subsection{Overtaking old positions}
We have just seen that Fermi's phenomenological approach resounded
in Majorana; in particular, it is through the latter's lectures on
Old Quantum Theory that Fermi's legacy is revealed. However, the
attention paid to phenomenology by a theoretician such as Majorana
should not be regarded as obvious. For example, we have already
mentioned that Dirac held that the educational approach to the new
physics should be ``essentially mathematical.''

In fact, Dirac and Fermi can be regarded as two very different
personifications of a theoretical physicist: more inclined to
mathematical formalization the former, to the use of intuitive
arguments the latter. These differences became conspicuous by
Fermi's well-known formulation of the quantum theory of radiation,
which he worked out between the end of 1920s and the beginning of
1930s since ``the method used by Dirac did not appeal to him''
\cite{amafnm305}. \footnote{Some placid rivalry possibly existed
between them, dating back to the formulation of the Fermi-Dirac
statistics \cite{rivalry}. E. Rutherford himself appears to make
fun of that when, in congratulating Fermi on his discovery of
neutron-induced radioactivity, he added: ``You may be interested
to hear that Professor Dirac also is doing some experiments. This
seems to be a good augury for the future of theoretical physics!''
\cite{ruther}.}

The attention to a phenomenological and intuitive approach
coexisted with a more mathematical formulation of quantum aspects
in Majorana's courses, and his teaching attitude may be regarded
as synthesizing Fermi's and Dirac's positions. Dirac was aware
that mathematics alone might hide the physical meaning of
theories, but simply he was not scared of it: ``One should learn
to hold the physical ideas in one's mind without reference to the
mathematical form.'' On the other hand, Majorana considered that
as a serious drawback: ``The mathematical method thoroughly
frustrates one's desire to intuit the physical meaning'' of the
formalism; not to mention that, according to him, the pre-existing
mathematical tools ``had been forced'' by Quantum Mechanics to
fulfill practical purposes.

Thus, an overtaking of Fermi's and Dirac's positions may be got
through Majorana's teaching attitude and, more explicitly, through
the prolusion to his course of 1938: the phenomenological
presentation of the Old Quantum Theory, borrowed from Fermi's
intuitive approach, was supplemented by the deepening of forefront
mathematical aspects of Quantum Mechanics. At the same time, the
limitations of a purely mathematical presentation - that sort of
presentation to which Dirac was inclined - were put forward by
Majorana, rising attention to more intuitive aspects connected
with quantum theories and their historical lines of development.

\section{Conclusions}

\noindent In the present paper, we suggested a comparison between
Fermi's and Majorana's attitudes towards the teaching of
Theoretical Physics. Fermi's 1927-28 course on Theoretical Physics
was crucial for our analysis. From Fermi's record book relative to
his 1927-28 course, a strong correspondence emerges between the
topics Fermi delivered in his lectures and six chapters of his
1928 book on Atomic Physics. This allowed a detailed
reconstruction of the content of his lectures on Theoretical
Physics, despite no written note being available on those
lectures.

The influence of Fermi's lectures can be perceived in Majorana's
personal notes dating back to the late 1920s. Further, Majorana's
personal notes and the programs of three courses he submitted in
Rome between 1933 and 1936 revealed his forefront approach and his
search for new routes in dealing with Quantum Mechanics.

We carried out a detailed comparison between Fermi's 1927-28
course on Theoretical Physics and the one Majorana gave in Naples
ten years later. On the basis of the available notes of Majorana's
course we showed that a number of Majorana's lectures - those
dealing with Old Quantum Theory - very closely reflect Fermi's
1927-28 course. That helped to conjecture about the content of
some of Majorana's previous lectures, for which no written notes
are available at all.

On the other hand, Fermi's approach just represented the starting
point for further theoretical developments in the reminder of
Majorana's course, devoted to the new Quantum Mechanics in both
its wave mechanical and matrix forms. Majorana's mathematical
approach, with the peculiar resort to Group Theory, proved his
original teaching touch. In fact, a thorough difference with
Fermi's lectures already emerged in the three programs submitted
by Majorana at the University of Rome. Majorana's courses thus
consisted in a fruitful mixture of an original approach - very
similar to that of the present courses in Quantum Mechanics - and
of some consolidated lines of development - which were the clear
legacy of the lines Fermi had adopted in his courses.

In conclusion, we have evidence of a deep influence of Fermi's
teaching approach on Majorana's one and, at the same time, a
thorough autonomy between their scientific personalities.

\newpage

\appendix

\section{Fermi's 1927-28 course on Theoretical Physics}

\noindent The topics treated by Fermi in his lectures (labelled by
their order number and date) are reported in the following for the
course in Theoretical Physics at the University of Rome in
1927-28.
\\ ${}$ \\
\begin{tabular}{lll}
N. & Date & Topics \\ \hline %
1 &(15 Nov)& Interactions between molecules. \\ & & Solids and
fluids in the framework of the kinetic theory.
\\
2 &(17 Nov)& Dependence of pressure upon molecular kinetic energy.
\\
3 &(19 Nov)& Mean free path. Vapor rays.
\\
4 &(22 Nov)& Kinetic energy in general coordinates. \\ & & The
theorem of equipartition of the energy (without proof).
\\
5 &(24 Nov)& Phase space.
\\
6 &(26 Nov)& Computation of the probability of a distribution.
\\
7 &(29 Nov)& The Boltzmann distribution law.
\\
8 &(1 Dec)& Proof of the theorem of equipartition of the energy.
\\
9 &(3 Dec)& The Maxwell law.
\\
10 &(6 Dec)& Principles of vector field theory.
\\
11 &(10 Dec)& Summary of the laws of electrology. Their
differential form.
\\
12 &(13 Dec)& Displacement currents. The Maxwell laws.
\\
13 &(15 Dec)& Propagation of plane electromagnetic perturbations.
\\ & & Their velocity.
\\
14 &(17 Dec)& Polarized sinusoidal plane wave.
\\
15 &(20 Dec)& Propagation of electromagnetic perturbations of
arbitrary shape.
\\
16 &(22 Dec)& Poynting vector.
\\
17 &(10 Jan)& Electromagnetic moment and radiation pressure.
\\
18 &(12 Jan)& Electronic theory of dispersion.
\\
19 &(14 Jan)& Electromagnetic radiation theory.
\\
20 &(17 Jan)& The electron.
\\
21 &(19 Jan)& Positive nuclei and the Rutherford atomic model.
\\
22 &(21 Jan)&  $\alpha$-, $\beta$-, and   $\gamma$-rays.
\\
23 &(24 Jan)& Radioactive transmutations.
\\
24 &(26 Jan)& Light quanta.
\\
25 &(28 Jan)& Compton effect.
\\
26 &(31 Jan)& Bohr's hypothesis.
\\
27 &(2 Feb)& Energy levels.
\\
28 &(4 Feb)& Kepler motions.
\\
29 &(7 Feb)& Dependence of the energy and the period on the
orbital \\ & & elements of motion.
\\
30 &(9 Feb)& Balmer series and the energy levels of hydrogen.
\\
31 &(11 Feb)& Orbits of the stationary states.
\\
32 &(14 Feb)& Determination of the Rydberg constant.
\\
33 &(25 Feb)& The spectrum of $He^+$ and the motion of the
nucleus.
\\
34 &(28 Feb)& Mechanical properties of the stationary orbits of
hydrogen.
\\
35 &(1 Mar)& Sommerfeld conditions for multi-periodic systems.
\\
36 &(6 Mar)& Application of the Sommerfeld conditions to the
oscillator \\ & & and the rotator.
\end{tabular}
\\ ${}$ \\
\begin{tabular}{lll}
N. & Date & Topics \\ \hline %
37 &(8 Mar)& Elliptic orbits of the hydrogen atom and hints on
Sommerfeld's \\ & & theory of the fine structure.
\\
38 &(10 Mar)& Proof of the relation dW/dJ = $\nu$  for a point
moving on a line. \\
39 &(13 Mar)& The correspondence principle for systems with one
\\ & & degree of freedom.
\\
40 &(15 Mar)& Harmonics of the motion for multi-periodic systems.
\\
41 &(17 Mar)& The correspondence principle for multi-periodic
systems.
\\
42 &(20 Mar)& Selection rules. Application to the azimuthal
quantum. \\ & & Evaluation of the mean-life of the hydrogen
states.
\\
43 &(22 Mar)& The principle of the adiabatic invariants.
\\ & & Relation between mechanical and magnetic moments. \\ & &
Bohr magneton.
\\
44 &(24 Mar)& Larmor theorem. Spatial quantization of the hydrogen
orbits.
\\
45 &(27 Mar)& The hypothesis of the spinning electron. \\ & &
Motion of the electronic axis and its spatial orientation.
\\
46 &(29 Mar)& Zeeman effect. Lorentz theory. \\ & & Quantum theory
of the hydrogen atom.
\\
47 &(31 Mar)& Hints on the Stark effect.
\\
48 &(17 Apr)& The electronic shielding.
\\
49 &(19 Apr)& Arc and spark spectrum.
\\
50 &(24 Apr)& Terms in a central, non Newtonian field.
\\
51 &(26 Apr)& Consequences of the spinning electron.
\\
52 &(28 Apr)& Determination of the weight of the orbits. \\ & &
The exclusion principle.
\\
53 &(1 May)& Stoner table. Chemical applications.
\\
54 &(3 May)& Applications to the alkali spectra.
\\
55 &(5 May)& Metals with two valence electrons.
\\
56 &(8 May)& Systems with singlets and triplets for the alkaline
earths.
\\
57 &(10 May)& Earth metals spectra.
\\
58 &(12 May)& Hints on the anomalous Zeeman effect. \\ & &
Anomalous Zeeman effect for the $D$ lines.
\\
59 &(15 May)& Hints on the Stark effect in heavy elements.
\\
60 &(19 May)& Review on the theory of gases.
\\
61 &(22 May)& Review on Boltzmann and Maxwell distribution laws.
\\
62 &(26 May)& Review on Electrodynamics.
\\
63 &(29 May)& Review on light quanta.
\\
64 &(31 May)& Review on the principle of the adiabatic invariants.
\\
65 &(2 Jun)&  Review on the correspondence principle.
\end{tabular}

\newpage

\section{Program of Majorana's first course}

\noindent {\bf Title:} Mathematical Methods of Quantum Mechanics \\
{\bf Academic Year:} 1933-34 \\

\noindent {\bf Topics to be covered:} \\
1) Unitary geometry. Linear transformations. Hermitian operators.
Unitary transformations. Eigenvalues and eigenvectors. \\
2) Phase space and the quantum of action. Modifications to
classical kinematics. General framework of Quantum Mechanics. \\
3) Hamiltonians which are invariant under a transformation group.
Transformations as complex quantities. Non compatible systems.
Representations of finite or continuous groups. \\
4) General elements on abstract groups. Representation theorems.
The group of spatial rotations. Symmetric groups of permutations
and other finite groups. \\
5) Properties of the systems endowed with spherical symmetry.
Orbital and intrinsic momenta. Theory of the rigid rotator. \\
6) Systems with identical particles. Fermi and Bose-Einstein
statistics. Symmetries of the eigenfunctions in the center-of-mass
frames. \\
7) The Lorentz group and the spinor calculus. Applications to
the relativistic theory of the elementary particles.\\
\noindent {\bf Date of submission:} May, 1933

\section{Program of Majorana's second course}

\noindent {\bf Title:} Mathematical methods of Atomic Physics \\
{\bf Academic Year:} 1935-36 \\

\noindent {\bf Topics to be covered:} \\
Matrix calculus. Phase space and the correspondence principle.
Minimal statistical sets or elementary cells. Elements of the
quantum dynamics.\\
Statistical theories. General definition of symmetry problems.
Representations of groups. Complex atomic spectra. Kinematics of
the rigid body. Diatomic and polyatomic molecules. \\
Relativistic theory of the electron and the foundations of
electrodynamics. \\
Hyperfine structures and alternating bands. Elements of Nuclear
Physics.\\
\noindent  {\bf Date of submission:} April 30, 1935

\section{Program of Majorana's third course}

\noindent {\bf Title:} Quantum electrodynamics \\
{\bf Academic Year:} 1936-37 \\

\noindent {\bf Topics to be covered:} \\
Relativistic theory of the electron. Quantization procedures.
Field quantities defined by commutability and anticommutability
laws. Their kinematical equivalence with sets with an undetermined
number of objects obeying the Bose-Einstein or Fermi statistics,
respectively. Dynamical equivalence. \\
Quantization of the Maxwell-Dirac equations. Study of the
relativistic invariance. The positive electron and the symmetry of
charges. \\
Several applications of the theory. Radiation and scattering
processes. Creation and annihilation of opposite charges.
Collisions of fast electrons.\\
\noindent {\bf Date of submission:} April 28, 1936

\[ \]


\end{document}